\documentclass[preprint,12pt]{elsarticle}
\usepackage{amssymb}

\begin{document}
\begin{frontmatter}

\title{Nonclassical properties of a particle in a finite range trap: the  $f$-deformed quantum oscillator approach}

\author[]{M. Davoudi Darareh \corref{cor1}}
\ead{m.davoudi@sci.ui.ac.ir}
\author[]{M. Bagheri Harouni}
\ead{m-bagheri@phys.ui.ac.ir}
\address{Department of Physics,
Faculty of Science, University of Isfahan, Hezar Jerib, Isfahan,
81746-73441, Iran} \cortext[cor1]{Corresponding author. Tel.:+98
311 7932435 ; fax: +98 311 7932409
 E-mail address: m.davoudi@sci.ui.ac.ir (M. Davoudi Darareh)}

\begin{abstract}
A particle bounded in a potential with finite range is described
by using an $f$-deformed quantum oscillator approach. Finite range
of this potential can be considered as a controllable deformation
parameter. The non-classical quantum statistical properties of
this deformed oscillator can be manipulated by nonlinearities
associated to the finite range.
\end{abstract}

\begin{keyword}
 Modified P\"{o}schl-Teller like coherent state \sep nonclassical property \sep
 $f$-deformed quantum oscillator
\PACS   03.65.Fd, 03.65.Ge, 42.50.Dv, 42.50.Ar
\end{keyword}

\end{frontmatter}

\section{Introduction}
The quantum harmonic oscillator,  its associated coherent states
and their generalizations~\cite{klauder skag perelomov ali} play
an important role in various theoretical and experimental fields
of modern physics, including quantum optics and atom optics.
Motivations for these generalizations have arisen from symmetry
considerations~\cite{klauder 1963}, dynamics~\cite{neito simmons
1978} and algebraic
aspects~\cite{jimbo,biden macfar}. \\
\indent The quantum groups approach~\cite{jimbo}  for generalizing
the notion of quantum harmonic oscillator and its realizations in
physical systems, by providing an algebraic method, has given the
possibility of extending the creation and annihilation operators
of the usual quantum oscillator to introduce the deformed
oscillator. In a very general important case, the associated
algebra of this deformed oscillator may be viewed as a
 deformation of classical Lie algebra by a generic function $f$,
the so-called  $f$-deformation function, depending nonlinearly on
the number of excitation quanta and some deformation parameters.
The corresponding oscillator is called an $f$-deformed
oscillator~\cite{manko 1 manko 55}. In contrast to the  usual
quantum harmonic oscillator,  $f$-deformed oscillators do not have
equally-spaced energy spectrum. Furthermore, it has been known
that the most of nonlinear generalizations of some physical
models, such as considered in ~\cite{crnug mar mik}, are only
particular cases of $f$-deformed models. Thus, it is reasonable
that $f$-deformed oscillators exhibit strongly various
nonclassical properties~\cite{manko 1 manko 55,roy,filho}, such as
the sub-Poissonian statistics, squeezing and the quantum
interference effects,  displaying the striking consequences of the
superposition principle of quantum mechanics. In addition,
$f$-deformed models depend on one or more deformation parameters
which should permit more flexibility and more ability for
manipulating  the model~\cite{davoudi,katriel solomon 49}. An
important question  in the $f$-deformed model is the physical
meaning of its deformation parameters. The $q$-deformed
oscillator~\cite{biden macfar}, as a special kind of $f$-deformed
oscillators with only one deformation parameter $q$, has been
extensively applied in describing  physical models, such as
vibrational and rotational spectra of molecules~\cite{chang yan
bonats argy ray bonatsos raychev rou smi}. The appearance of
various nonclassical features induced by a $q$-deformation
relevant to some specific nonlinearity is also
studied~\cite{katriel solomon
artoni zang birman}. \\
\indent Based on the above-mentioned considerations,  $f$-deformed
 quantum oscillators and their associated coherent states, such as
 $f$-coherent states ~\cite{manko 1 manko 55} or nonlinear coherent
 states~\cite{filho}, can be appropriately established in
 attempting to describe certain physical phenomena where their
 effects could be modelled through  a deformation on  their
 dynamical algebra with respect to conventional or usual
 counterparts. This approach has been accomplished, for instance, in the
 study of the stationary states of the center-of-mass motion of an ion in the harmonic trap~\cite{filho} and under
 effects associated with the curvature of physical space~\cite{mahdifar
 vogel}, the influence of the spatial confinement  on  the
 center-of-mass motion of an exciton in a quantum  dot~\cite{bageri
 41}, the influence of atomic collisions and the finite number of
 atoms in a Bose-Einstein condensate on controlled manipulation of
 the nonclassical properties of radiation field~\cite{davoudi},
 some nonlinear processes in high intensity photon beam~\cite{manko 1 manko
 55}, intensity-dependent atom-field interaction in absence and in presence  of nonlinear quantum
 dissipation in a micromaser~\cite{naderi 32 naderi 39} and finally, incorporating the effects of interactions among
 the particles in the framework of the $q$-deformed algebra~\cite{scarfone
 41}. \\
 \indent It is shown that the trapped systems provide a powerful tool
 for  preparation and manipulation of
 nonclassical states~\cite{liebfried}, quantum
 computations~\cite{bennett} and quantum communications~\cite{braunstein
 loock}. Improved experimental techniques have caused precise measurements on realistic trapping systems, for example,
  trapped  ion-laser systems~\cite{meekhof monroe}, trapped  gas of
  atoms~\cite{anderson}  and electron-hole carriers confined in a
  quantum well and  quantum dot~\cite{harrison}. A study of
  confined quantum systems using the Wood-Saxon
  potential~\cite{costa} and the  $q$-analogue harmonic oscillator
  trap~\cite{sharma sharma}, are some  efforts which can be used to explain some
  experimentally observed deviations from the results predicted by
  calculations based on the harmonic oscillator model. \\
  \indent A realistic case in any experimental setup is that the
  dimension of the trap is finite  and the realistic trapping
  potential is not the harmonic oscillator potential extending to
  infinity. Thus, the realistic confining potential becomes flat near the edges of the trap and
  can be simulated by the  tanh-shaped potential  $V(x)=D\,tanh^2(x/\delta)$,
  so-called the modified (or hyperbolic) P\"{o}schl-Teller(MPT)
  potential~\cite{mpt potential}. The MPT potential presents
  discrete (or bound) and continuum (or scattering) states. The
  dynamical symmetry algebra associated with the bound part of the
  spectrum is $su(2)$ algebra ~\cite{frank isacker} while for the
  complete spectra is $su(1,1)$ algebra ~\cite{arias gomez lemus}.
   The MPT potential has been used very widely  in many branches of
   physics, such as, atom optics~\cite{wang},  molecular physics~\cite{frank wolf lemus bernal} and
    nanostructure   physics~\cite{harrison}.\\
   \indent Constructing coherent states for systems with discrete and continuous spectrum~\cite{gazeau klauder} and for various
   kinds of confining potentials~\cite{antoine gazeau klauder} have become a very important tool in the
   study of some quantum systems. The P\"{o}schl-Teller(PT) potentials,
    including trigonometric PT(TPT) and MPT potentials with
    discrete  infinite and finite dimensional bound states respectively, because of their relations to several
 other trapping potentials are of crucial importance.
        Some types of the coherent states for the MPT potential have been constructed. The minimum-uncertainty coherent states
   formalism ~\cite{neito 2}, the Klauder-Perelomov
   approach~\cite{klauder skag perelomov ali} by realization of
   lowering and raising operators in terms of the physical variable $u=tanh(x/\delta)$ by means of
   factorizations~\cite{cruz kuru negro} and applying one kind
   generalized deformed oscillator algebra with a selected deformed commutation relation~\cite{daskaloyannis
   91}, are some attempts for this purpose. \\
   \indent In the  present paper, we intend to investigate the
   nonlinear effects appeared due to finite dimension of the
   trapping potential on producing new nonclassical quantum
   statistical properties  using the
   $f$-deformed quantum oscillator approach. For this aim, it
   will be shown that the finite range of the trapping potential
   leads to the  $f$-deformation of the  usual harmonic potential with the well  depth  $D$
   as a controllable physical deformation
   parameter. Then, the $f$-deformed  bound  coherent states~\cite{recamier
   jauregui} for the above-mentioned MPT quantum oscillator are
   introduced and their nonclassical properties are examined. We think that by this $f$-deformed quantum oscillator approach
   the problem of trapped ion-laser system and trapped gas of atoms, such as a  Bose-Einstein condensate, in a realistic trap
   can be studied analytically.  \\
   \indent The paper is organized as follows. In section \ref{f deformed}, we
   introduce the $f$-deformed quantum oscillator equivalent to the
   MPT oscillator  and obtain the associated ladder operators. In
   section \ref{f coherent states}, we construct the $f$-deformed
   bound coherent states of the MPT quantum oscillator and examine
   its  resolution of identity. Section \ref{statistics} devoted to the study of the influence of the
   finite range potential on producing and manipulating the
   nonclassical properties, including the sub-Poissonian statistics and squeezing character. Finally, the summary and conclusions are presented in
    section \ref{conclusion}.

\section{MPT Hamiltonian as an $f$-deformed quantum
oscillator}\label{f deformed}
 In this section, we will consider a
bounded particle inside the MPT potential, called  the MPT
oscillator, and we will associate to this system an $f$-deformed
quantum oscillator. By using this mathematical model, we try to
investigate physical deformation parameters in the model, to
manipulate the nonlinearities related to the finite range effects
on this system. For this purpose, we first give the bound energy
eigenvalues for the MPT potential. Then, by comparing it with the
energy spectrum of the general $f$-deformed quantum oscillator, we
will obtain the deformed annihilation and
creation operators.\\
\indent Let us consider the MPT potential energy

\begin{equation}\label{mpt potential}
V(x)=D\,tanh^2(\frac{x}{\delta}),
\end{equation}
where $D$ is the depth of the well, $\delta$ determines the range
of the potential and $x$ gives the relative distance from the
equilibrium position. The well depth, D,  can  be defined as
$D=\frac{1}{2}m\omega^2\delta^2$, with mass of the particle $m$
and angular frequency $\omega$ of the harmonic oscillator, so
that, in the limiting case $D\rightarrow \infty$(or $
\delta\rightarrow \infty)$, but keeping the product $m\omega^2$
finite, the MPT potential energy reduces to harmonic potential
energy,  $\lim_{D\rightarrow \infty}
V(x)=\frac{1}{2}m\omega^2x^2$. Figure 1 depicts the MPT  potential
for three different values of the well depth $D$. Harmonic
potential limit by increasing $D$ is clear from this figure.
Solving the Schr\"{o}dinger equation, the energy eigenvalues for
the MPT potential are obtained as~\cite{landau}

\begin{equation}\label{mpt energy 1}
E_n=D-\frac{\hbar^2\omega^2}{4D}(s-n)^2,     \quad \quad n=0, 1,
2, \cdots, [s]
\end{equation}
in which $s=(\sqrt{1+(\frac{4D}{\hbar\omega})^2}-1)/2$,  and $[s]$
stands for the closest integer to $s$ that is smaller than $s$.
The MPT oscillator quantum number $n$ can not be larger than the
maximum number of bound states $[s]$, because of the dissociation
 condition $s-n\geq 0$. Consequently, the total number of bound
 states is $[s]+1$. We should note that for integer $s$, the  final bound
 state and the total number of bound states will be $s-1$ and $s$,
 respectively. Also, for every small value of the well depth $D$,
 we always have at least one bound state for the MPT oscillator, i.e., the ground  state. By introducing a dimensionless    parameter
  $N=\frac{4D}{\hbar\omega}=\frac{2m\omega\delta^2}{\hbar}$, the total number of bound
  states will obtain from $[(\sqrt{1+(N)^2}-1)/2]+1$. For
  integer $s$, a simple relation $N=2\sqrt{s(s+1)}$ will connect $N$
  to the total number of bound  states, i.e., $s$. The
  bound energy spectrum in equation (\ref{mpt energy 1}) can be
  rewritten as

\begin{equation}\label{mpt energy 2}
E_n=\hbar\omega[-\frac{n^2}{N}+(\sqrt{1+\frac{1}{N^2}}-\frac{1}{N})n+\frac{1}{2}(\sqrt{1+\frac{1}{N^2}}-\frac{1}{N})].
\end{equation}
The relation (\ref{mpt energy 2}) shows a nonlinear dependence on
the quantum number $n$, so that, different energy levels are not
equally spaced. It is clear that, in the limit $D\rightarrow
\infty$ (or $N\rightarrow \infty$), the energy spectrum for the
quantum  harmonic oscillator will be obtained, i.e.,
$E_n=\hbar\omega(n+\frac{1}{2})$. In  contrast with some confined
systems such as a particle bounded in an infinite and finite
square well potentials, by decreasing the size of the confinement
parameter, i.e., the finite range $\delta$ of the MPT oscillator,
energy eigenvalues decreases.\\ \indent A quantity that has a
close connection   to experimental information is the energy level
spacing, $E_{n+1}-E_{n}$, where it corresponds to the transition
frequency between two adjacent energy levels. Furthermore, by this
quantity one  can theoretically explore an algebraic
representation for the quantum mechanical potentials with discrete
spectrum~\cite{wunsche}. Based upon above considerations, a useful
illustration for the effects of the deformation parameter $D$ on
the nonlinear behavior of the deformed oscillator, can be
investigated by introducing the delta parameter $\Delta_n$ as

\begin{equation}\label{delta param}
\Delta_n=\frac{E_{n+1}-E_{n}}{\hbar\omega}-1
\end{equation}

\noindent which measures the amount of deviation of the adjacent
energy level spacing  of the deformed oscillator with respect to
the non-deformed or harmonic oscillator. Substituting from
equation (\ref{mpt energy 2}) in equation (\ref{delta param}) we
can ontain the delta parameter $\delta$ for the MPT potential
\begin{equation}\label{delta param 2}
\Delta_n=-\frac{2}{N}n+\sqrt{1+\frac{1}{N^2}}-\frac{2}{N}-1
\end{equation}
 \indent On the other
hand, the $f$-deformed quantum oscillator \cite{manko 1 manko 55},
as a nonlinear oscillator with a specific kind of nonlinearity, is
characterized by the following deformed dynamical variables
$\hat{A}$ and  $\hat{A}^\dag$

\begin{eqnarray}\label{fd}
      \hat{A}&=&\hat{a}f(\hat{n})=f(\hat{n}+1)\hat{a},\nonumber\\
   \hat{A}^\dag&=&f(\hat{n})\hat{a}^\dag=\hat{a}^\dag f(\hat{n}+1),     \quad
         \quad
         \hat{n}=
         \hat{a}^\dag\hat{a},
    \end{eqnarray}

   \noindent where $\hat{a}$ and  $\hat{a}^\dag$ are usual boson
    annihilation and creation operators $([\hat{a},
    \hat{a}^\dag]=1)$, respectively. The real deformation function
    $f(\hat{n})$ is a nonlinear operator-valued function of the
    harmonic number operator $\hat{n}$, where it introduces some
    nonlinearities to the system. From equation (\ref{fd}), it
    follows that the $f$-deformed operators $\hat{A}$,
    $\hat{A}^\dag$ and $\hat{n}$ satisfy the following closed
    algebra
    \begin{eqnarray}\label{algebrafd}
     &[\hat{A}, \hat{A}^\dag]=&(\hat{n}+1)f^2(\hat{n}+1)-\hat{n}f^2(\hat{n}),\nonumber\\
    &[\hat{n}, \hat{A}]=&-\hat{A},       \quad      \quad
    [\hat{n}, \hat{A}^{\dag}]=\hat{A}^{\dag}.
     \end{eqnarray}

     \noindent The above-mentioned  algebra, represents a deformed
     Heisenberg-Weyl algebra whose nature depends on the nonlinear
     deformation function $f(\hat{n})$.
     An $f$-deformed oscillator is a nonlinear  system
     characterized by a Hamiltonian of the harmonic oscillator
     form

     \begin{equation}\label{hamiltf}
     \hat{H}=\frac{\hbar\omega}{2}(\hat{A}^\dag\hat{A}+\hat{A}\hat{A}^\dag).
     \end{equation}

     Using equation (\ref{fd}) and the number state representation
     $\hat{n}|n \rangle=n|n \rangle$, the eigenvalues of the
     Hamiltonian (\ref{hamiltf}) can be written as

     \begin{equation}\label{energyf}
     E_n=\frac{\hbar\omega}{2}[(n+1)f^2(n+1)+nf^2(n)].
     \end{equation}
     \indent It is worth  noting that in the limiting  case $f(n)\rightarrow 1$, the deformed
     algebra (\ref{algebrafd}) and  the deformed energy eigenvalues
     (\ref{energyf}) will
     reduce to the conventional Heisenberg-Weyl
     algebra and the harmonic oscillator spectrum, respectively.\\
     \indent Comparing the bound energy spectrum of the MPT oscillator,
     equation (\ref{mpt energy 2}), and the energy spectrum of an
     $f$-deformed oscillator, equation (\ref{energyf}), we obtain
     the corresponding deformation function for the MPT oscillator as

     \begin{equation}\label{f}
     f^2(\hat{n})=\sqrt{1+\frac{1}{N^2}}-\frac{\hat{n}}{N}.
     \end{equation}

     Furthermore, the ladder operators of the bound eigenstates of
     the MPT Hamiltonian can be written in terms of the
     conventional operators $\hat{a}$ and $\hat{a}^\dag$ as
     follows

     \begin{equation}\label{a mpt}
     \hat{A}=\hat{a}\sqrt{\sqrt{1+\frac{1}{N^2}}-\frac{\hat{n}}{N}},
     \quad    \quad
     \hat{A}^\dag=\sqrt{\sqrt{1+\frac{1}{N^2}}-\frac{\hat{n}}{N}}\hat{a}^\dag.
     \end{equation}

     \noindent These two operators satisfy the deformed Heisenberg-Weyl commutation relation

     \begin{equation}\label{algebra mpt}
     [\hat{A},
     \hat{A}^\dag]=\sqrt{1+\frac{1}{N^2}}-\frac{2\hat{n}+1}{N},
     \end{equation}

     \noindent and they act upon the quantum number states $|n\rangle$, corresponding to the
     energy eigenvalues $E_n$  given in equation (\ref{mpt energy
     2}), as

     \begin{eqnarray}\label{def ladder}
     \hat{A}|n\rangle&=&f(n)\sqrt{n}|n-1 \rangle, \nonumber\\
     \hat{A}^{\dag}|n\rangle&=&f(n+1)\sqrt{n+1}|n+1 \rangle.
     \end{eqnarray}

     \indent The commutation relation (\ref{algebra mpt}), can be
     identified with the usual $su(2)$ commutation relations by
     introducing the set of transformations

     \begin{equation}\label{transform j}
     \hat{A} \rightarrow  \frac{\hat{J}_+}{\sqrt{N}}, \quad
     \hat{A}^{\dag} \rightarrow  \frac{\hat{J}_-}{\sqrt{N}}, \quad
     \hat{n} \rightarrow  \frac{\sqrt{1+N^2}-1}{2}-\hat{J}_0,
     \end{equation}

     \noindent where $\hat{J}_\mu$ satisfy the usual angular momentum
     relations ~\cite{rose}. The $f$-deformed commutation relation
     (\ref{algebra mpt}) in a special case of large but finite value of $N$, which corresponds to the small deformation,
     can lead to a maths-type $q$-deformed commutation
     relation~\cite{arik}, i.e., $\hat{A}\hat{A}^{\dag}-q\hat{A}^{\dag}\hat{A}=1$, with
     $q=1-\frac{2}{N}=1-\frac{\hbar\omega}{2D}$. The harmonic oscillator
     limit corresponds to $D\rightarrow \infty$ then $q\rightarrow
     1$. This result confirms a correspondence between the
     $q$-deformed oscillators and finite range potentials, which
     is studied elsewhere~\cite{balles civ reb monat das kok}.\\
     \indent It is evident that, herein, we have focused our attention on the quantum states
     of the MPT Hamiltonian which exhibit bound oscillations with
     finite range. The remaining states, i.e., the scattering states or energy continuum eigenstates, have
      non-evident boundary conditions. From physical point of
      view, it means that the excitation energies of this confined
      system in the MPT potential energy are small compared with
      the well depth potential energy $D$, such that, only the
      vibrational modes dominated and the scattering or continuum
      states should be neglected. Some important physical systems
      with such circumstances are vibrational excitations of
      molecular systems \cite{carrington choi moore},
      trapped ions or atoms~\cite{song hai luo} and the electron-hole
      carriers confined in a quantum well~\cite{harrison}.\\

       \section{$f$-Deformed bound coherent states}\label{f coherent states}

       In the context of the $f$-deformed quantum oscillator
       approach, we introduce the $f$-deformed bound coherent states $|\alpha,f\rangle$ for the MPT
       oscillator as a coherent superposition of  all bound energy
       eigenstates of the MPT Hamiltonian as below

       \begin{equation}\label{alfa f}
       |\alpha,f\rangle=C_f\sum_{n=0}^{[s]}\frac{\alpha^n}{\sqrt{n!}f(n)!}|n\rangle,
       \quad
       C_f=\left(\sum_{n=0}^{[s]}\frac{|\alpha|^{2n}}{n!(f(n)!)^2}\right)^{-1/2},
       \end{equation}

       \noindent so that $\hat{n}|n\rangle=n|n\rangle$,  and $f(n)!=f(n)f(n-1)\cdots f(0)$, where $f(n)$ is
       obtained in equation (\ref{f}). Since the sum in the
       equation (\ref{alfa f}) is finite, the states $|\alpha,f\rangle$,
       similar to the Klauder-Perelomov coherent states ~\cite{klauder skag perelomov ali}, are not an eigenstate of the
       annihilation operator $\hat{A}$. From equations  (\ref{def ladder})
       and  (\ref{alfa f}), we arrive at
       \begin{equation}\label{a alfa f}
       \hat{A}|\alpha,f\rangle=\alpha|\alpha,f\rangle-\frac{C_f\alpha^{[s]+1}}{\sqrt{[s]!}f([s])!}|[s]\rangle.
       \end{equation}
       As is clear from this equation, these states can not be considered as a right-hand eigenstate of annihilation
       operator $\hat A$. This property is common character of all coherent states that are defined in a finite-
       dimensional basis~\cite{recamier jauregui,buzek 1}.\\
       \indent The ensemble of the $f$-deformed bound coherent states
       $|\alpha,f\rangle$ labelled by the complex number $\alpha$
       form an overcomplete set with the resolution of the identity

       \begin{equation}\label{resolution}
     \int d^2\alpha|\alpha,f\rangle
     m_f(|\alpha|)\langle\alpha,f|=\sum_{n=0}^{[s]}
     |n\rangle\langle n|=\hat{\textbf{1}},
     \end{equation}

     \noindent where $m_f(|\alpha|)$ is the proper measure for this family
     of the  bound coherent states. Substituting from equation (\ref{alfa f}) in equation
     (\ref{resolution}) and using integral relation $\int _{0}^{\infty}K_\nu (t)t^{\mu-1}dt=
     2^{\mu-2}\Gamma(\frac{\mu-\nu}{2})\Gamma(\frac{\mu+\nu}{2})$
     for the modified Bessel function $K_\nu (t)$ of the second kind and of the
     order $\nu$,  we obtain the suitable choice for the measure
     function as

      \begin{equation}\label{measure}
     m_f(|\alpha|)=\frac{K_\nu
     (|\alpha|)}{2^l \pi |\alpha|^\nu C_f^2(|\alpha|)},
     \end{equation}

     \noindent where $\nu=(1+\gamma)n-\eta$, $l=(1-\gamma)n+\eta+1$ and
     $\gamma=\frac{1}{N}$, $\eta=\sqrt{1+\frac{1}{N^2}}$.

     \indent In contrast to the Gazeau-Klauder coherent
     states~\cite{gazeau klauder}, the $f$-deformed  coherent
     states, such as introduced in equation (\ref{alfa f}), do not generally
     have the temporal stability~\cite{manko 1 manko 55}. But it
     is possible to introduce a notion of temporally stable
     $f$-deformed coherent states~\cite{roknizadeh tavassoly}.

       \section{Quantum statistical properties of the MPT
       oscillator}\label{statistics}
       \subsection{Sub-Poissonian statistics}

       In order to determine the quantum statistics of the MPT quantum oscillator, we consider
        Mandel parameter $Q$ defined by~\cite{mandel}

        \begin{equation}\label{mandel param1}
     Q=\frac{\langle\hat{n}^2\rangle-\langle\hat{n}\rangle^2}{\langle\hat{n}\rangle}-1.
     \end{equation}

     The sub-Poissonian statistics (antibunching effect), as an
     important nonclassical property, exists whenever $Q <0$. When
     $Q >0$, the state of the system is called super-Poissonian (bunching
     effect). The state with $Q =0$ is called Poissonian.
     Calculating the Mandel parameter $Q$ in equation (\ref{mandel
     param1}) over the $f$-deformed bound coherent states $|\alpha, f\rangle$ defined in
     equation (\ref{alfa f}), it can be described the
     finite range dependence of the Mandel parameter. Figure 2 shows the  parameter $Q$ for
     four different values of $|\alpha|$, i.e., $|\alpha|=3,\,
     4,\,5,\,7$. As is seen, for every one of the  values of $|\alpha|$, the Mandel parameter $Q$ exhibits the
     sub-Poissonian statistics  at certain range of $D$ or the dimensionless
     parameter $N=\frac{4D}{\hbar\omega}$, where this range is
     determined by  the value of  $|\alpha|$. The bigger
     parameter $|\alpha|$ is, the more late the Mandel parameter
     tends to the Poissonian statistics. As expected, with further increasing values
     of $D$ or $N$, the Mandel parameter $Q$ finally stabilized at
     an asymptotical zero value, corresponding to the Poissonian
     statistics associated to the canonical harmonic oscillator
     coherent states. For the limit $N\rightarrow
     0$(or $D\rightarrow 0$) and for every values of $|\alpha|$,
     the Mandel parameter becomes $Q=-1$, where it is reasonable,
     because in this limit, only  the ground state supports by the potential.
     \subsection{Quadrature squeezing}
     As another important nonclassical property, we examine the
     quadrature squeezing of the MPT quantum oscillator. For this
     purpose, we  consider quadrature operators $\hat{q}_{\varphi}$ and
     $\hat{p}_{\varphi}$ defined as ~\cite{dodonov}

     \begin{equation}\label{quadr operat}
     \hat{q}_{\varphi}=\frac{1}{\sqrt{2}}(\hat{a}e^{-i\varphi}+\hat{a}^\dag
     e^{i\varphi}), \quad    \quad
     \hat{p}_{\varphi}=\frac{i}{\sqrt{2}}(\hat{a}^{\dag}e^{i\varphi}-\hat{a}e^{-i\varphi}),
     \end{equation}

     \noindent satisfying the commutation relation
     $[\hat{q}_{\varphi},\hat{p}_{\varphi}]=i$. One can define the
     invariant squeezing coefficient $S$ as the difference between
     the minimal value (with respect to the phase $\varphi$) of
     the variances of each quadratures and the mean value $1/2$ of
     these variances in the coherent or vacuum state. Simple
     calculations result in the formula

     \begin{equation}\label{}
     S=\langle\hat a^{\dag}\hat a\rangle-|\langle \hat
     a\rangle|^2-|\langle\hat a^2 \rangle-\langle\hat a\rangle^2|,
     \end{equation}

     \noindent so that the condition of squeezing is $S<0$.
     Calculating the squeezing parameter $S$
        over the $f$-deformed bound coherent states in equation
        (\ref{alfa f}), we examine the squeezed character of these states. In figure 3, we
     have plotted the parameter $S$ with respect to the dimensionless
     deformation parameter $N=\frac{4D}{\hbar\omega}$ for three
     different values of $|\alpha|$, namely  $|\alpha|=0.5,\, 1,\,
     1.3$. As is seen, the states $|\alpha, f\rangle$ exhibit squeezing for
     certain values of $|\alpha|$. Furthermore, the
     squeezing character of the states $|\alpha, f\rangle$  tend
     to zero as $N$ or the well depth $D$ of the MPT potential
     approaches to infinity, according to the coherent states of
     the quantum harmonic oscillator. In the limit $N\rightarrow
     0$(or $D\rightarrow 0$), this plot shows the quadrature
     squeezing $S=0$, where it is in agreement with the  only ground
      state supported by the potential in this limit.

     \section{Conlusions}\label{conclusion}

     In this paper, we have introduced an algebraic approach based
     on the $f$-deformed quantum oscillator for considering a
     particle in the real confining potential which has finite trap
     dimension, in contrast  to the harmonic oscillator potential
     extending to infinity. Proposed confining model potential is
     the modified P\"{o}schl-Teller potential. We have shown that
     the effects of the finite trap dimension in this model
     potential  can be considered as a natural deformation in
     the quantum harmonic oscillator algebra. This quantum
     deformation approach makes possible analytical study of a wide category of
     realistic bound quantum systems algebraically. It is shown
     that the nonlinear behavior resulted from this finite range
     effects can lead to generate and  manipulate some important nonclassical
     properties for this deformed quantum oscillator. We have
     obtained that the presented $f$-deformed bound coherent
     states of the modified P\"{o}schl-Teller potential can
     exhibit the sub-Poissonian statistics and quadrature
     squeezing in definite domain of the trap dimension or well depth $D$ of this potential.
      In the large but finite value for the well depth $D$,
     i.e., small deformation, a $q$-deformed oscillator with $q=1-\hbar\omega/(2D)$ will
     result. In the limit $D\rightarrow \infty$, the harmonic
     oscillator counterpart is obtained.\\ \indent
     Based on the
     approach in this paper, we can obtain exact solutions for
     realistic confined physical systems such as, trapped ion-laser system (in
     progress), Bose-Einstein condensate and confined carriers in
     nano-structures.\\ \\
       {\bf Acknowledgments}\\
       The authors wish to thank The Office of Graduate Studies and Research Vice President of The University of Isfahan for their support.

\newpage

{\bf Figure captions}\\

Fig. 1. Plots of the MPT potential for three different values of
the well depth $D$, $D=1$(solid curve), $D=2$(dashed curve) and
$D\rightarrow \infty$(dotted curve).\vspace{1cm}

Fig. 2. Plots of the Mandel parameter $Q$ versus the dimensionless
deformation parameter $N$ for $|\alpha|=3$(solid curve),
$|\alpha|=4$(dashed curve), $|\alpha|=5$(dotted curve) and
$|\alpha|=7$(dash-dotted curve). \vspace{1cm}

Fig. 3. Plots of the invariant squeezing coefficient $S$ versus
the dimensionless deformation parameter $N$ for
$|\alpha|=0.5$(solid curve), $|\alpha|=1$(dashed curve) and
$|\alpha|=1.3$(dotted curve).

\end{document}